\documentclass[useAMS,usenatbib]{mn2e}

\usepackage{graphicx}
\usepackage[fleqn]{amsmath}
\usepackage{amssymb}
\usepackage{color}

\newcommand{\cs}{c_\mathrm{s}}

\newcommand{\betac}{\beta_\mathrm{c}}

\title[Stochastic disc fragmentation]{Numerical convergence in self-gravitating shearing sheet simulations and the stochastic nature of disc fragmentation}

\author[S.-J. Paardekooper]{Sijme-Jan Paardekooper$^1$\thanks{E-mail: 
S.Paardekooper@damtp.cam.ac.uk}
\\  
$^1$DAMTP, University of Cambridge, Wilberforce Road, Cambridge CB3 0WA,
}

\begin{document}

\date{Draft version \today}

\pagerange{\pageref{firstpage}--\pageref{lastpage}} \pubyear{2011}

\maketitle

\label{firstpage}

\begin{abstract}
We study numerical convergence in local two-dimensional hydrodynamical simulations of self-gravitating accretion discs with a simple cooling law. It is well-known that there exists a steady gravito-turbulent state, in which cooling is balanced by dissipation of weak shocks, with a net outward transport of angular momentum. Previous results indicated that if cooling is too fast (typical time scale 3 $\Omega^{-1}$, where $\Omega$ is the local angular velocity), this steady state can not be maintained and the disc will fragment into gravitationally bound clumps. We show that, in the two-dimensional local approximation, this result is in fact not converged with respect to numerical resolution and longer time integration. Irrespective of the cooling time scale, gravito-turbulence consists of density waves as well as transient clumps. These clumps will contract because of the imposed cooling, and collapse into bound objects if they can survive for long enough. Since heating by shocks is very local, the destruction of clumps is a stochastic process. High numerical resolution and long integration times are needed to capture this behaviour. We have observed fragmentation for cooling times up to $20$ $\Omega^{-1}$, almost a factor $7$ higher than in previous simulations. Fully three-dimensional simulations with a more realistic cooling prescription are necessary to determine the effects of the use of the two-dimensional approximation and a simple cooling law.
\end{abstract}
 


\section{Introduction}
The notion that gaseous planets may form by gravitational instability (GI) dates back to \cite{kuiper51} and \cite{cameron78}. Although revived by \cite{boss97}, the disc instability model suffered a blow when it was realised that very efficient cooling was needed for it to operate \citep{gammie01}. This makes GI ineffective in planet forming regions (1 - 20 AU) of typical protoplanetary discs \citep[e.g.][]{rafikov05,matzner05}. The recent discovery of planets orbiting at very large distances from their central star \citep{kalas08, la10}, where cooling time scales are sufficiently short, has sparked new interest in the possibility of planet formation via disc instability \citep[e.g.][]{boley09}. For a review of gravitational instability and its applications in planet formation theory, see \cite{durisen07}.

A razor-thin self-gravitating disc is unstable to axisymmetric perturbations when \citep{safronov60,toomre64}:
\begin{equation}
Q\equiv \frac{\cs \kappa}{\pi G\Sigma} < 1,
\label{eqQ}
\end{equation} 
where $\cs$ is the sound speed, $\kappa$ is the epicyclic frequency, $G$ is the gravitational constant and $\Sigma$ is the surface density. For Keplerian discs, $\kappa$ equals the angular velocity of the disc $\Omega$. Discs are in fact unstable to non-axisymmetric perturbations at slightly higher values of $Q$ \citep{papa91}. Dissipation of the resulting spiral waves provides a source of heating, which can balance radiative cooling. An equilibrium can then be set up \citep{pac78}, in which gravitational and Reynolds stresses are such that the energy losses due to cooling are compensated for by dissipation. If we parametrise the cooling time scale as
\begin{equation}
t_\mathrm{cool}=\beta \Omega^{-1},
\label{eqtcool}
\end{equation}
then in equilibrium we must have that \citep{pringle81,gammie01}:
\begin{equation}
\alpha = \frac{4}{9}\frac{1}{\gamma(\gamma-1)\beta},
\label{eqalpha}
\end{equation}
where $\alpha$ is the usual stress parametrisation \citep{shakura78} and $\gamma$ is the ratio of specific heats. This equilibrium of gravito-turbulence can therefore lead to significant transport of angular momentum in regions of the disc where cooling is efficient and $Q \sim 1$. 

The above parametrisation assumes angular momentum transport is local, i.e. that the spiral waves damp close to their location of excitation \citep{cossins09,forgan11}. In principle, there can be a non-local part in the transport \citep{balbus99}, but Smoothed Particle Hydrodynamics (SPH) simulations find that transport is essentially local for low disc masses $M_\mathrm{disc}/M_* < 0.25$ \citep{lodato04,lodato05}. In the present work, we restrict ourselves to \emph{local} simulations, for which the boundary conditions ensure local transport \citep{balbus99}. 

Naturally, the critical value for $t_\mathrm{cool}$, which can be translated into a critical value of $\beta$, $\betac$, through equation (\ref{eqtcool}), below which fragmentation is inevitable, has received much attention, since this defines the region in discs where planets can form by GI. \cite{gammie01} found that $\betac=3$ for 2D local simulations with $\gamma=2$. \cite{rice05} interpreted the critical value for $\beta$ as a maximum value of $\alpha$ the disc can sustain (see equation (\ref{eqalpha})). They found $\alpha_\mathrm{max}=0.06$, which is in agreement with the $\betac$ found by \cite{gammie01}. Note that this implies that $\betac$ depends on $\gamma$.

The simple parametrisation of the cooling time scale in terms of the local dynamical time scale is of course an oversimplification. There have been several attempts to improve on the simple cooling law of equation (\ref{eqtcool}). \cite{nelson00} discuss an implementation of photospheric cooling, while the simulations of \cite{boss01} and \cite{boley06} include full radiative transfer in their global models. \cite{johnson03} pointed out that care has to be taken in applying the simple cooling criterion $\beta < \betac$ to discs with realistic cooling. The development of non-linear structures in the disc can cause the cooling time scale to change significantly from that of the initial state. More recently, \cite{kratter11} and \cite{rice11} studied the effect of irradiation by the central star on disc fragmentation. It is generally agreed that for these more complicated models, similar to the case of a simple cooling law, fast cooling is needed to trigger fragmentation. 

Recently, numerical convergence of the critical cooling time scale was questioned in \cite{meru11}, who found fragmentation at much higher values of $\beta$ than previously found at lower resolution. It was pointed out in \cite{conv} that this behaviour can at least partly be explained by the effect of starting from smooth initial conditions. In global simulations with constant $\beta$, where the cooling time scale therefore depends on radius through $\Omega$ in equation (\ref{eqtcool}), the inner parts of the disc get turbulent before the outer parts. The edge between the turbulent and non-turbulent region of the disc can trigger fragmentation at longer cooling times. In local simulations, both $\beta$ and $t_\mathrm{cool}$ are constant, and therefore no edges appear. However, even in this case one has to worry about initial conditions: if the time scale to set up gravito-turbulence is longer than the cooling time scale, the disc will quickly cool down to $Q < 1$, which triggers the strong linear instability, possibly leading again to artificial fragmentation. 

The influence of artificial viscosity on fragmentation in SPH simulations was discussed in \cite{lodato11}, who suggested that artificial heating could play a major role in fragmentation even if it constitutes only $5\%$ of the total heating. \cite{pickett07} showed that in locally isothermal, grid-based three-dimensional simulations, artificial viscosity affects the survival of clumps. It is clear that disc fragmentation is a difficult problem to handle numerically, and therefore it makes sense to go back to the simplest possible set-up in which fragmentation can be studied, namely, the two-dimensional shearing sheet.

Numerical convergence in local simulations with respect to $\betac$ has not been thoroughly discussed so far. \cite{gammie01} showed convergence with respect to the value of $\alpha$ measured in the simulations, but it remains to be seen if the fragmentation criterion is independent of numerical resolution. This is the subject of the present work. The paper is organised as follows: in Sect. \ref{secEq}, we present the basic equations in the framework of the shearing sheet, and outline the numerical method used to solve these equations in Sect. \ref{secNum}. We discuss the performance of the numerical method on two test problems in Sect. \ref{secTest}. In Sect. \ref{secInit}, we discuss the initial conditions used, and present the results in Sect. \ref{secRes}. We discuss the results in Sect. \ref{secDisc} and conclude in Sect. \ref{secCon}.
  
\section{Basic equations}
\label{secEq}
Following \cite{gammie01}, we use the shearing sheet approximation \citep{gold65}, in which we follow a patch of the disc centred at cylindrical coordinates $(R,\varphi)=(R_0,\varphi_0+\Omega t)$, where $\Omega$ is the angular velocity at $R=R_0$. On this patch, we define a Cartesian coordinate frame defined by $x=R-R_0$ and $y=R_0(\varphi-\varphi_0-\Omega t)$, and expand the equations of motion to first order in $|x|/R_0$. This leads to a linear shear in the patch. The basic equations then are the continuity equation:
\begin{equation}
\frac{\partial \Sigma}{\partial t}+\nabla \cdot \Sigma {\bf v}=0,
\end{equation}
with $\Sigma$ the surface density and ${\bf v}=(v_x,v_y)^T$ the 2D velocity vector, Euler's equation:
\begin{equation}
\frac{\partial {\bf v}}{\partial t}+{\bf v}\cdot \nabla {\bf v} + \frac{\nabla p}{\Sigma} = 2q\Omega^2 x {\bf \hat x}-2\Omega \times {\bf v}-\nabla \Phi,
\end{equation}
where $p$ is the 2D pressure, $q$ is the shear parameter ($q=3/2$ in a Keplerian disc), $\Omega$ is the angular velocity of the coordinate frame, and $\Phi$ is the self-gravity potential. Finally, we have the equation for internal energy density:
\begin{equation}
\frac{\partial \epsilon}{\partial t}+\nabla \cdot \epsilon {\bf v} = -p \nabla \cdot {\bf v}-\frac{\epsilon\Omega}{\beta},
\end{equation}
where the last term represents our simple cooling law. We will assume a perfect gas throughout so that $p=(\gamma-1)\epsilon$.

The self-gravity potential of a three-dimensional density distribution $\rho$ can be found from
\begin{equation}
\Phi({\bf x})=-G\int \frac{\rho({\bf x'}) d^3{\bf x'}}{|{\bf x}-{\bf x'}|}.
\end{equation}
Taking $\rho = \Sigma(x,y)\delta(z-\delta)$, we get
\begin{equation}
\Phi(x,y,z)=-G\int \frac{\Sigma(x',y') dx dy}{\sqrt{(x-x')^2+(y-y')^2+(z-\delta)^2}},
\end{equation}
from which we see that $\Phi(z=0)$ gives the potential due to a surface density $\Sigma$, but smoothed over a length $\delta$. Such a potential would arise when integrating the three-dimensional equations in the vertical direction over the thickness of the disc. 

For a surface density that is periodic in all directions and for a single Fourier component ${\bf k}=(k_x,k_y)^T$, the above equation has the solution
\begin{equation}
\Phi_{\bf k} = -2\pi G \frac{\Sigma_{\bf k}}{|{\bf k}|}\exp(-|{\bf k}|\delta).
\end{equation}
The smoothing affects Fourier components on scales smaller than $\delta$. Note that both \cite{gammie01} and \cite{rice11} use $\delta=0$, allowing for small scales in the potential (up to the grid scale).
 

We assume a background state that has constant surface density $\Sigma_0$ and sound speed $c_\mathrm{s,0}$. This defines both a $Q$-value for the patch (equation (\ref{eqQ})) and an approximate measure of the disc thickness $H=c_\mathrm{s,0}/\Omega$. Note that in the presence of self-gravity, $H$ is no longer exactly the pressure scale height. 

It is convenient to split-off the Keplerian part of the velocity and work with the perturbed velocity ${\bf u} = (v_x,v_y+q\Omega x)^T$:
\begin{equation}
\frac{\partial \Sigma}{\partial t}-q\Omega x\frac{\partial\Sigma}{\partial y}+\nabla \cdot \Sigma {\bf u}=0,
\end{equation}
\begin{equation}
\frac{\partial {\bf u}}{\partial t}-q\Omega x\frac{\partial{\bf u}}{\partial y}+{\bf u}\cdot \nabla {\bf u} + \frac{\nabla p}{\Sigma} = q\Omega v_x {\bf \hat y}-2\Omega \times {\bf u}-\nabla \Phi,
\end{equation}
\begin{equation}
\frac{\partial \epsilon}{\partial t}-q\Omega x\frac{\partial \epsilon}{\partial y}+\nabla \cdot \epsilon {\bf u} = -p \nabla \cdot {\bf u}-\frac{\epsilon\Omega}{\beta}.
\label{eqeint}
\end{equation}
Expressing these in conservation form, defining momenta $m=\Sigma v_x$ and $n=\Sigma (v_y+q\Omega x)$, we get:
\begin{equation}
\frac{\partial \Sigma}{\partial t}-q\Omega x\frac{\partial \Sigma}{\partial y}+\frac{\partial m}{\partial x}+\frac{\partial n}{\partial y}=0,
\label{eqcont}
\end{equation}
\begin{eqnarray}
\frac{\partial m}{\partial t}-q\Omega x\frac{\partial m}{\partial y}+\frac{\partial}{\partial x}\left(\frac{m^2}{\Sigma} + p\right)+\frac{\partial}{\partial y}\left(\frac{mn}{\Sigma}\right)=\nonumber\\
2\Omega n -\Sigma\frac{\partial\Phi}{\partial x},
\label{eqmomx}
\end{eqnarray}
\begin{eqnarray}
\frac{\partial n}{\partial t}-q\Omega x\frac{\partial n}{\partial y}+\frac{\partial}{\partial x}\left(\frac{mn}{\Sigma}\right)+\frac{\partial}{\partial y}\left(\frac{n^2}{\Sigma}  +p\right) =\nonumber\\
 (q-2)\Omega m-\Sigma\frac{\partial\Phi}{\partial y},
 \label{eqmomy}
\end{eqnarray}
\begin{eqnarray}
\frac{\partial e}{\partial t}-q\Omega x\frac{\partial e}{\partial y}+\frac{\partial}{\partial x}\left((e+p)\frac{m}{\Sigma}\right)+\frac{\partial}{\partial y}\left((e+p)\frac{n}{\Sigma}\right)=\nonumber\\ 
q\Omega \frac{mn}{\Sigma} -m\frac{\partial\Phi}{\partial x}-n\frac{\partial\Phi}{\partial y}-\frac{p\Omega}{(\gamma-1)\beta},
\label{eqenergy}
\end{eqnarray}
where $e=(m^2+n^2)/2\Sigma + \epsilon$ is the total energy density associated with the velocity perturbations. Note that we have removed the contribution from the shear from the total energy. Together with taking no background gradients, this ensures that all quantities $\Sigma$, $m$, $n$, $p$, $\Phi$ and $e$ are shear-periodic. Splitting off the background shear comes at the expense of an extra source term in the energy equation, which represents the work done by Reynolds stress due to the background shear \citep[first term on the right-hand side of equation (\ref{eqenergy}), see also][]{stone10}.

\section{Numerical method}
\label{secNum}
Since we are dealing with numerical convergence, we go into some detail explaining the numerical method used. We use operator splitting to solve the different parts of equations (\ref{eqcont}) - (\ref{eqenergy}). Since we are interested in a balance between cooling and shock heating, it makes sense to use a Riemann solver to deal with the hydrodynamics. Throughout, we will use a uniform Cartesian grid covering the shearing sheet of size $L_x\times L_y$, with grid spacings $\Delta x$ and $\Delta y$. 

\subsection{Time step}
For stability, the time step is limited by the Courant-Friedrichs-Lewy (CFL) condition \citep{courant}:
\begin{equation}
\Delta t = C_0 \min\left(\frac{\Delta x}{|u|+\cs}, \frac{\Delta y}{|v|+\cs}\right),
\end{equation}
where $u=m/\Sigma$, $v=n/\Sigma$, and the minimum is taken over all grid cells and $C_0$ is the Courant number. We have used $C_0=0.4$ throughout. Note that it is the \emph{perturbed} velocity that enters the time step calculation. Because the large background shear velocity has been split off, much larger time steps are possible \citep{masset00a}. In low resolution studies, care must be taken that the time step does not become so large that two neighbouring rows are sheared apart \citep{masset00b}. This is never an issue for the resolutions adopted in this paper. 

\subsection{Orbital advection}
\label{secOrb}
Given a time step $\Delta t$, the most straightforward integration step involves just the first two terms of equations (\ref{eqcont}) - (\ref{eqenergy}), which is linear advection at the background Keplerian shear \citep{masset00a,masset00b}. In the following, we adopt the usual convention that $X^n_{i,j}$ denotes quantity $X$ evaluated at time index $n$ and space indices $i$ (in the $x$ direction) and $j$ (in the $y$ direction). For each row at $x_i$, we therefore need to shift the solution by $\delta y = q\Omega x_i \Delta t$. This can be achieved for quantity $X$ by first shifting the solution by an integer number of grid cells $N$, where $N$ is the nearest integer to $\delta y/\Delta y$. The rest of the shift $\delta y-N\Delta y$ can be done to second order accuracy by standard methods \citep[e.g.][]{leveque}:
\begin{equation}
X^{n+1}_{i,j} = X^n_{i,j} - |a|(X^n_{i,j} - X^n_{i,u}) - \frac{1}{2}a(1-|a|)(\sigma_{i,j} - \sigma_{i,u}),  
\label{eqshift} 
\end{equation} 
where $a=\delta y/\Delta y - N$, $\sigma$ denotes a limited slope, and $u=j-a/|a|$ is the upwind direction. The second term on the right-hand side gives a first-order update, while the third term provides second-order corrections. Note that since $N$ is the nearest integer to $\delta y/\Delta y$, we always have that $|a| < 1/2$, so that the above scheme is stable for any value of $\Delta t$. For the slope limiter, we use the same form as used in the Riemann solver (superbee, see Sect. \ref{secRoe}). 
  
\cite{stone10} include the extra energy source term (first term on the right-hand side in equation (\ref{eqenergy})) in the orbital advection step. We found this can occasionally lead to negative pressures, and that stability is improved by instead integrating this source term during the $x$ integration (see below). Furthermore, it proved numerically advantageous to shift the pressure according to equation (\ref{eqshift}) rather than the total energy, again to avoid unphysical states when $e$ is dominated by kinetic energy.
   
\subsection{Cooling}
It is numerically convenient to split off the non-geometrical source terms and integrate them separately \citep{eulderink95}. In our case, this involves only the cooling source term, so in this step we solve
\begin{equation}
\frac{\partial e}{\partial t}=-\frac{p\Omega}{(\gamma-1)\beta},
\end{equation}
or, since all other state variables ($\Sigma$, $m$ and $n$) are constant during this step:
\begin{equation}
\frac{\partial p}{\partial t}=-\frac{p\Omega}{\beta}.
\label{eqpcool}
\end{equation} 
For the cases we are interested in, $\Delta t \ll \beta/\Omega$, and we have found that a first order integration of equation (\ref{eqpcool}) gave good enough results. 
   
 \subsection{Gravitational potential}
 \label{secPot}
To calculate the gravitational potential $\Phi$, we basically follow \cite{gammie01}. We first shift back the density in $y$ to the time it was last periodic ($t=t_p$). For this, we use the same algorithm as used for orbital advection. The resulting surface density is completely periodic in $x$ and $y$, and the potential can then be found from
\begin{equation}
\Phi = -2\pi G \sum_{\bf k} \frac{\Sigma_{\bf k}}{|{\bf k}|}\exp (i{\bf k}\cdot {\bf x}-|{\bf k}|\delta),
\end{equation}
where $\Sigma_{\bf k}$ are the Fourier components of the shifted $\Sigma$ and ${\bf k}=(k_x+q\Omega (t-t_p) k_y, k_y)^T$. This calculation can be done most effectively by using the Fast Fourier Transform (FFT). First, the Fourier components of the surface density are calculated using the FFT, the result of which is divided by $|{\bf k}|$. An inverse FFT then gives us $\Phi$. This potential is then shifted forward in $y$ to the present time, after which the forces are obtained by taking finite differences. Even when using $\delta=0$, the use of finite differences naturally introduces a smoothing length comparable to the grid scale for the gravitational force. 

\subsection{Integrating in the $x$ direction}
Taking only terms involving derivatives with respect to $t$ and $x$ from equations (\ref{eqcont}) - (\ref{eqenergy}), together with the appropriate geometrical source terms \citep{eulderink95}, we get:
\begin{equation}
\frac{\partial \Sigma}{\partial t}+\frac{\partial m}{\partial x}=0,
\label{eqcont2}
\end{equation}
\begin{equation}
\frac{\partial m}{\partial t}+\frac{\partial}{\partial x}\left(\frac{m^2}{\Sigma} + p\right) = 2\Omega n - \Sigma\frac{\partial \Phi}{\partial x},
\end{equation}
\begin{equation}
\frac{\partial n}{\partial t}+\frac{\partial}{\partial x}\left(\frac{mn}{\Sigma}\right)= \left(q-2\right)\Omega m,
\end{equation}
\begin{equation}
\frac{\partial e}{\partial t}+\frac{\partial}{\partial x}\left((e+p)\frac{m}{\Sigma}\right)=q\Omega \frac{mn}{\Sigma}-m\frac{\partial\Phi}{\partial x}.
\label{eqenergy2}
\end{equation}
Note that all source terms in the energy equation are due to the kinetic part of the total energy: the thermal energy does not change directly due to work done by self-gravity or the Reynolds stress due to the background shear (see also equation (\ref{eqeint})). Note also that this would not hold if we had integrated the energy source term associated with the background shear during the orbital advection step as in \cite{stone10}. 

\subsubsection{Roe solver}
\label{secRoe}
Ignoring the source terms for the moment (we come back to these in section \ref{secSource} below), the left-hand sides of the above equations have exactly the structure of ordinary Cartesian hydrodynamics, even though we are working with \emph{fluctuations} on top of the background shear. Therefore, we can use standard techniques to deal with these. As mentioned above, since we are interested in shock heating, it makes sense to use a Riemann solver. This is advantageous not only when dealing with shocks, but for waves in general, because of the use of a characteristic decomposition of fluctuating quantities. 

The above equations (still ignoring source terms) can be written concisely as $\partial{\bf W}/\partial t + \partial {\bf F}/\partial x=0$, where ${\bf W} = (\Sigma, m, n, e)^T$ and ${\bf F}=(m, m^2/\Sigma+p, mn/\Sigma, (e+p)m/\Sigma)^T$. Recalling that $u=m/\Sigma$, the Jacobian matrix $A=d{\bf F}/d{\bf W}$ has eigenvalues $u+\cs$, $u-\cs$ and $u$, and eigenvectors 
\begin{eqnarray}
{\bf e}_1 &=& (1, u+\cs, v, h+\cs u)^T,\label{eqe1}\\
{\bf e}_2 &=& (1,u-\cs,v, h-\cs u)^T,\\
{\bf e}_3 &=& (0,0,1,v)^T,\\
{\bf e}_4 &=& (1,u,v,\frac{u^2}{2}+\frac{v^2}{2})^T\label{eqe4},
\end{eqnarray}
where $v=n/\Sigma$ and
\begin{equation}
h=\frac{u^2}{2}+\frac{\omega^2}{2}+\frac{\gamma}{\gamma-1}\frac{p}{\Sigma}.
\end{equation}

A vector ${\bf \Delta}=(\Delta_\Sigma, \Delta_m, \Delta_n, \Delta_e)^T$ can be projected onto the eigenvectors of $A$ using the coefficients
\begin{eqnarray}
\lefteqn{a_1 =\frac{\gamma-1}{2 \cs^2}\left[\left(\frac{u^2}{2}+\frac{v^2}{2}\right)\Delta_\rho  +
\right.} \nonumber\\
&&\left. \frac{ }{ } \Delta_e-u\Delta_m - v\Delta_n \right]+
\frac{\Delta_m - u\Delta_\rho}{2\cs}\label{eqa1}\\
\lefteqn{a_2=\frac{\gamma-1}{2 \cs^2}\left[\left(\frac{u^2}{2}+\frac{v^2}{2}\right)\Delta_\rho  +\right.} \nonumber\\
&&\left. \frac{ }{ } \Delta_e-u\Delta_m - v\Delta_n \right]-\frac{\Delta_m - u\Delta_\rho}{2\cs}\\
\lefteqn{a_3=\Delta_n - v\Delta_\rho}\\
\lefteqn{a_4=\frac{\gamma-1}{\cs^2}\left[\left(h-u^2-v^2\right)\Delta_\rho  - \Delta_e+u\Delta_m + v\Delta_n \right].}\label{eqa4}
\end{eqnarray}

A \emph{linear} hyperbolic system of the form $\partial {\bf W}/\partial t+A\partial {\bf W}/\partial x=0$ can be solved by projecting the state difference between neighbouring grid cells onto the eigenvectors of $A$, or, in other words, decomposing the state difference into characteristic waves:
\begin{equation}
{\bf W}_i-{\bf W}_{i-1} = \sum_k a_k {\bf e}_k,
\end{equation}
where we have omitted the subscript $j$ for clarity. The state at the cell interface is found by subtracting all right-going waves from ${\bf W}_i$:
\begin{equation}
{\bf W}_{i-1/2} = {\bf W}_i - \sum_k \max(\mathrm{sign}(\lambda_k),0) a_k {\bf e}_k,
\end{equation}
where $\lambda_k$ is the $k$th eigenvalue of $A$. The interface flux ${\bf F}_{i-1/2}=A{\bf W}_{i-1/2}$:
\begin{equation}
{\bf F}_{i-1/2}= {\bf F}_i - \sum_k \frac{\mathrm{sign}(\lambda_k)+1}{2} a_k\lambda_k {\bf e}_k.
\end{equation}
Alternatively, the interface state can be found by taking into account all left-going waves from ${\bf W}_{i-1}$:
\begin{equation}
{\bf W}_{i-1/2} = {\bf W}_{i-1} - \sum_k \min(\mathrm{sign}(\lambda_k),0) a_k {\bf e}_k,
\end{equation}
from which we can again obtain an interface flux
\begin{equation}
{\bf F}_{i-1/2}= {\bf F}_{i-1} - \sum_k \frac{\mathrm{sign}(\lambda_k)-1}{2} a_k\lambda_k {\bf e}_k.
\end{equation}
Combining both expressions for ${\bf F}_{i-1/2}$, we get

\begin{equation}
{\bf F}_{i-1/2} = \frac{1}{2}\left( {\bf F}_{i-1} + {\bf F}_i - \sum_k |\lambda_k| a_k {\bf e}_k\right),
\label{eqFlux}
\end{equation}
A first-order state update is then given by
\begin{equation}
{\bf W}_i^{n+1}={\bf W}_i^n +\frac{\Delta t}{\Delta x}\left( {\bf F}_{i-1/2} - {\bf F}_{i+1/2}\right).
\label{eqUpdate}
\end{equation}

Equations (\ref{eqcont2})-(\ref{eqenergy2}) are in fact \emph{non-linear}. However, \cite{roe81} introduced a suitable linearisation, in which the matrix $A$ is replaced by a matrix that is averaged over two neighbouring grid cells. By demanding that the resulting solution should be exact if the two cells were connected by a single shock wave, \cite{roe81} found that the eigenvectors (\ref{eqe1})-(\ref{eqe4}) and projection coefficients (\ref{eqa1})-(\ref{eqa4}) should be evaluated at the so-called Roe-averaged state
\begin{eqnarray}
\hat u = \frac{\sqrt{\Sigma_i} u_i + \sqrt{\Sigma_{i-1}} u_{i-1}}{\sqrt{\Sigma_i} + \sqrt{\Sigma_{i-1}}},\\
\hat v = \frac{\sqrt{\Sigma_i} v_i + \sqrt{\Sigma_{i-1}} v_{i-1}}{\sqrt{\Sigma_i} + \sqrt{\Sigma_{i-1}}},\\
\hat h = \frac{\sqrt{\Sigma_i} h_i + \sqrt{\Sigma_{i-1}} h_{i-1}}{\sqrt{\Sigma_i} + \sqrt{\Sigma_{i-1}}},
\end{eqnarray}
and an average sound speed 
\begin{equation}
\hat{c}_\mathrm{s}^2 = (\gamma-1)\left(\hat h - \frac{1}{2} \hat u^2 - \frac{1}{2} \hat v^2\right).
\end{equation}
We can then use the flux function (\ref{eqFlux}) to update the state using (\ref{eqUpdate}). 

One can obtain a method that is second-order accurate in both space and time by using modified projection coefficients \citep{eulderink95,leveque}
\begin{equation}
\tilde a_k = a_k\left(1 - \frac{\Delta t}{\Delta x} |\lambda_k|\right) .
\end{equation}
However, near discontinuities any method that is more than first-order accurate will introduce unphysical oscillations \citep{godunov}. It is therefore necessary to introduce a flux limiter $\phi$:
\begin{equation}
\tilde a_k = a_k\left(1 - \phi(r_k) \frac{\Delta t}{\Delta x} |\lambda_k|\right),
\end{equation}
where $r_k = a_k / a_{ku}$ is a parameter comparing $a_k$ to the value of $a_k$ in the upwind direction. In smooth flow, we expect $r_k \approx 1$ and we obtain a second-order method if $\phi(1)=1$. The functional form of $\phi$ is chosen so that no oscillations are introduced by the term proportional to $\phi$, which can be made mathematically precise using the concept of Total Variation (TV) \citep[e.g.][]{leveque}. Any chosen limiter function must by TV \emph{Diminishing} (TVD) in order for it not to introduce oscillations near shocks. A fairly general class of TVD limiters can we written as
\begin{equation}
\phi (\theta)=\max (0,\min (1,s\theta), \min (s,\theta)),
\label{eqfluxlimiter}
\end{equation}
which is TVD for $1 \leq s \leq 2$. The case $s=1$ corresponds to the minmod limiter, which is the most diffusive choice, while $s=2$ corresponds to the superbee limiter, which in general gives the sharpest shocks. Because of the use of a flux limiter, no explicit artificial viscosity is needed to stabilise the numerical scheme. This however comes at the price of giving up second-order accuracy in non-smooth regions of the flow. In all simulations presented, we have used the superbee flux limiter. 
   
\subsubsection{Source terms}
\label{secSource}
The system $\partial {\bf W}/\partial t + \partial {\bf F}/\partial x = {\bf S}$ can be solved by combining the Riemann solver solution to $\partial {\bf W}/\partial t + \partial {\bf F}/\partial x = 0$ with either a solution to $d{\bf W}/dt = {\bf S}$ \emph{or} $d{\bf F}/dx = {\bf S}$. The latter choice, called stationary extrapolation \citep{eulderink95}, is the preferred choice when a balance between source terms and flux gradients is expected to arise in the solution. This is the case for example in global simulations of Keplerian discs \citep{rodeo}, where in the radial direction there exists a balance between gravity, the centrifugal force and pressure. In the present case, we are dealing directly with perturbations. It is then more advantageous to use $d{\bf W}/dt = {\bf S}$.

In particular, it is possible to deal with epicyclic oscillations in a way as to conserve the energy associated with epicyclic motion to round-off error \citep{gardiner05,gressel07,stone10}. Recall that we are solving 
\begin{eqnarray}
\frac{dm}{dt}=2\Omega n + S_x, \label{eqSx}\\
\frac{dn}{dt}=(q-2)\Omega m\label{eqSy},
\end{eqnarray}
where $S_x=-\Sigma \partial\Phi/\partial x$ is the source term due to self-gravity, together with $d\Sigma/dt = dp/dt=0$. The energy associated with epicyclic motions can be conserved by integrating these equations using a Crank-Nicholson scheme \citep{stone10}:
\begin{eqnarray}
\left[m\right]&=&2\Delta t\Omega {\bar{n}}+\Delta t \bar{S_x}, \\
\left[n\right]&=&\left(q-2\right)\Delta t\Omega {\bar{m}},
\end{eqnarray}
where $\bar x = (x^{n+1}+x^n)/2$ is a time average, and $[x]$ is short for $x^{n+1}-x^n$. Solving these two equations gives:
\begin{eqnarray}
\left[m\right]=\frac{2\Delta t\Omega\left(2 n^n - \Delta t \Omega \left(2-q\right)m^n+\bar{S_x}/\Omega\right)}{2+\Delta t^2 \Omega^2  \left(2-q\right)},\label{eqCNx}\\
\left[n\right]=-2\Delta t \Omega \left(2-q\right)\frac{m^n+\Delta t \bar{S_x}/2+\Delta t \Omega n^n}{2+\Delta t^2 \Omega^2  \left(2-q\right)}\label{eqCNy}.
\end{eqnarray}
Note that $S_x = S_x(\Sigma)$, which is therefore constant during the source term integration. 

As an alternative, $S_x$ could be left out in this step, and integrated using stationary extrapolation \citep{eulderink95}. This could be advantageous when objects form in which pressure is balanced by self-gravity. This balance would then be recognised by the Roe solver, resulting in no evolution away from this steady state. 

\subsubsection{A full time step}
\label{secFullTimeStep}
A complete integration step for the $x$ direction then consists of the following steps:
\begin{itemize}
\item{Prediction: evolve the momenta under influence of the source terms using equations (\ref{eqSx}) and (\ref{eqSy}) for half a time step. A simple Euler integration is sufficient in this step.}
\item{Riemann solver: use the output of the previous step to calculate a state update ${\bf dW} _\mathrm{hydro}$ for a full time step.}
\item{Source integration: use equations (\ref{eqCNx}) and (\ref{eqCNy}) calculate the source update to the state ${\bf dW}_\mathrm{source}$, keeping $\Sigma$ and $p$ constant. For the state variables on the right-hand side we use ${\bf W}^n + {\bf dW}_\mathrm{hydro}/2$. Note that we need $S_x$ for both $\Sigma^n$ and $\Sigma^{n+1}$, so that we need to recalculate the gravitational potential after the previous step.}
\item{Update the state according to ${\bf W}^{n+1}={\bf W}^n + {\bf dW}_\mathrm{source} + {\bf dW}_\mathrm{hydro}$.}
\end{itemize}

\subsection{Integrating in the $y$ direction}
For the $y$-direction, we proceed in a similar way. The governing equations in terms of $m$ and $n$ read:
\begin{equation}
\frac{\partial\Sigma}{\partial t} +\frac{\partial n}{\partial y}=0,
\end{equation}
\begin{equation}
\frac{\partial m}{\partial t} +\frac{\partial}{\partial y}\left(\frac{mn}{\Sigma}\right)=0,
\end{equation}
\begin{equation}
\frac{\partial n}{\partial t}+\frac{\partial}{\partial y}\left(\frac{n^2}{\Sigma}+p\right)=-\Sigma\frac{\partial \Phi}{\partial y},
\end{equation}
\begin{equation}
\frac{\partial e}{\partial t} +\frac{\partial}{\partial y}\left((e+p)\frac{n}{\Sigma}\right)=-n\frac{\partial\Phi}{\partial y}.
\end{equation}

\subsubsection{Riemann solver}
The left-hand sides of above equations have the same structure as their $x$-equivalents, which leads to similar expressions for the eigenvectors and projection coefficients:
\begin{eqnarray}
{\bf e}_1 &=& (1, u, v+\cs, h+\cs v)^T,\\
{\bf e}_2 &=& (1,u, v-\cs, h-\cs v)^T,\\
{\bf e}_3 &=& (0,1,0,u)^T,\\
{\bf e}_4 &=& (1,u,v,\frac{u^2}{2}+\frac{v^2}{2})^T,
\end{eqnarray}
and 
\begin{eqnarray}
\lefteqn{a_1 =\frac{\gamma-1}{2 \cs^2}\left[\left(\frac{u^2}{2}+\frac{v^2}{2}\right)\Delta_\rho  +
\right.} \nonumber\\
&&\left. \frac{ }{ } \Delta_e-u\Delta_m - v\Delta_n \right]+
\frac{\Delta_n - v\Delta_\rho}{2\cs}\\
\lefteqn{a_2=\frac{\gamma-1}{2 \cs^2}\left[\left(\frac{u^2}{2}+\frac{v^2}{2}\right)\Delta_\rho  +\right.} \nonumber\\
&&\left. \frac{ }{ } \Delta_e-u\Delta_m - v\Delta_n \right]-\frac{\Delta_n - v\Delta_\rho}{2\cs}\\
\lefteqn{a_3=\Delta_m - u\Delta_\rho}\\
\lefteqn{a_4=\frac{\gamma-1}{\cs^2}\left[\left(h-u^2-v^2\right)\Delta_\rho  - \Delta_e+u\Delta_m + v\Delta_n \right].}
\end{eqnarray}
Again, using Roe-averaging and flux-limiting, these formulae can be used to construct a second-order correct update for the state vector. 

\subsubsection{Source term integration}
There are no epicyclic oscillations when considering the $y$-direction only, and we can therefore adopt a simple integration scheme
\begin{equation}
n^{n+1}=n^n + \Delta t \bar{S_y},
\end{equation}
where $S_y= - \Sigma \partial \Phi/\partial y$, together with $d\Sigma/dt = dm/dt=dp/dt=0$. Alternatively, stationary extrapolation could be used for this step. The complete integration step is equivalent to that in the $x$-direction (see section \ref{secFullTimeStep}).
 
\subsection{Backup fluxes}
We expect strong density and pressure contrasts to arise in the simulations, especially in cases that fragmentation will happen. These conditions provide a challenge for numerical methods, and it is important to assess where the method could fail. Since Riemann solvers require the use of the total energy, they are more likely to fail (i.e. predict negative pressures) when $\left|{\bf u}\right| \gg \cs$ \citep{einfeldt91}.

The maximum possible time step restricted by the CFL condition is $\Omega \Delta t \sim \Omega \Delta x / \cs = \Delta x/ H$. Provided that $H$ is properly resolved (as it should be), we have that $\Omega t \ll 1$. For $\beta \geq 1$, we therefore expect the simple Euler integration of the cooling not to give unphysical states. 

The source term integrations during the $x$ and $y$ integration steps all involve $d\Sigma/dt = dp/dt=0$. This means that the states fed into the Riemann solver by the prediction step (see section \ref{secFullTimeStep}) are all physical (i.e. $\Sigma > 0$ and $p>0$). The only place where the scheme can break down is therefore the Riemann solver itself giving negative pressures or surface densities. This can for example happen when the Riemann problem between two cells is not linearisable \citep{einfeldt91}.

A possible way of dealing with such failures is to add more diffusion by returning to the first-order 
fluxes locally. Although the first order Roe fluxes are readily available, they can still produce unphysical states if the Riemann problem can not be linearised. As an alternative, we use the more diffusive, Harten-Lax-van Leer (HLL) solver \citep{hll}. This scheme is in fact positive definite \citep{einfeldt91}, which means that, given physical input states, it will never lead to negative pressures or densities. The corresponding flux function is, in the $x$-direction, given by 
\begin{equation}
{\bf F}_{i-1/2}^\mathrm{HLL}=\frac{b^+ {\bf F}_{i-1} - b^- {\bf F}_{i} + b^- b^+ ({\bf W}_{i} - {\bf W}_{i-1})}{b^+ - b^-},
\end{equation}
where 
\begin{eqnarray}
b^- &=& \min_k(0,\hat\lambda_{k}, \lambda_{k,i-1}),\\
b^+&=&\max_k(0,\hat\lambda_{k}, \lambda_{k,i})
\end{eqnarray} 
are measures of the minimum and maximum possible wave speeds encountered in the Riemann problem. Here, $\lambda_{k,i}$ are the eigenvalues of the Jacobian matrix (see Sect. \ref{secRoe}) based on the state and flux of cell $i$, and $\hat\lambda_k$ are the eigenvalues based on the Roe-averaged state between cells $i$ and $i-1$. If the Roe flux is found to lead to an unphysical state, the flux is replaced by the HLL flux. Usually, this is only necessary once in every $10^7$ updates.

\subsection{Boundary conditions}
Periodic boundary conditions are used in the $y$ direction. In the $x$ direction, the sheet is shear-periodic; for example at the inner boundary we have that
\begin{equation}
\Sigma(x,y)=\Sigma(x+L_x,y-q\Omega L_x t),
\end{equation}
where $L_x$ is the size of the sheet in the $x$ direction. The required shift in $y$ is again performed using the same method as used for orbital advection (see Sect. \ref{secOrb}). 

\section{Test Problems}
\label{secTest}
The Riemann solvers were tested using standard one-dimensional shock tubes and two-dimensional Riemann problems. Below, we describe two test problems that are specific to the shearing sheet.

\begin{figure} 
\centering
\resizebox{\hsize}{!}{\includegraphics[]{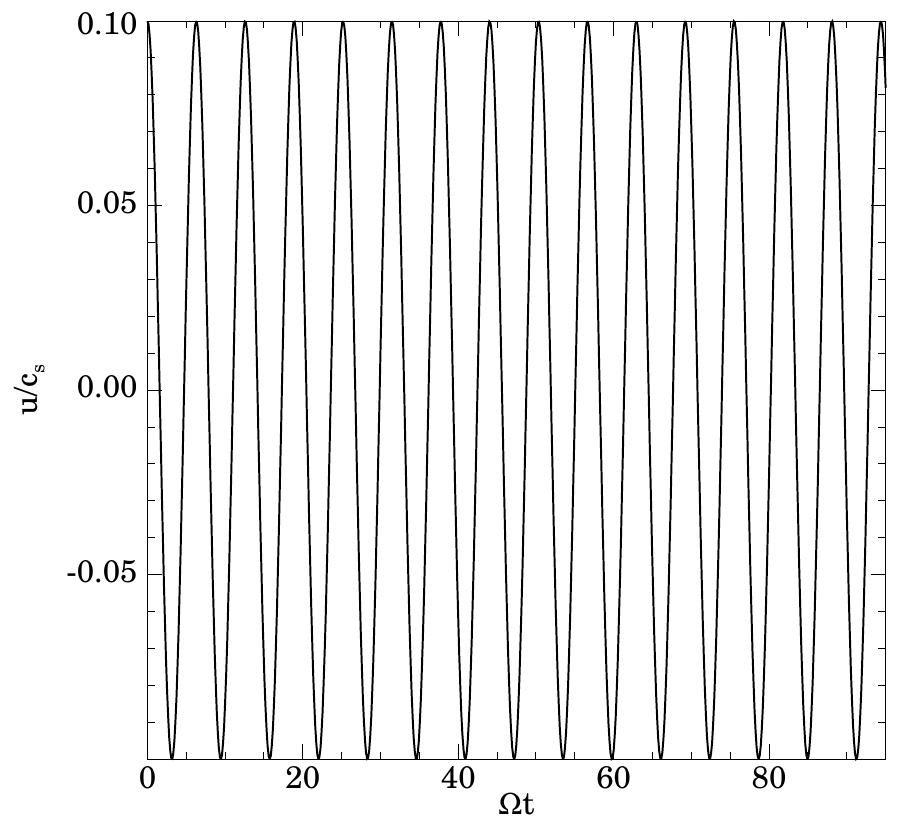}} 
\caption{Epicyclic motion in the absence of any gradients on a grid of $128^2$, $-1/2 \leq (x,y) \leq 1/2$, and $\cs=0.01$.} 
\label{figepi} 
\end{figure} 

\subsection{Epicyclic motion}
In the absence of self-gravity and any gradients in $x$ or $y$, there exist oscillating solutions to the governing equations:
\begin{eqnarray}
m(t)=m_0 \cos \kappa t +\frac{2\Omega n_0}{\kappa} \sin \kappa t,\\
n(t)=n_0\cos\kappa t - \frac{\kappa m_0}{2\Omega} \sin\kappa t,
\end{eqnarray} 
where $\kappa^2=2(2-q)\Omega^2$ is the square of the epicyclic frequency. In a Keplerian disc, with $q=3/2$, we have that $\kappa=\Omega$. 

We take a grid of $128^2$, $-1/2 \leq (x,y) \leq 1/2$, take $\cs=0.01$ and give the whole grid a velocity perturbation of $u=0.1\cs$. The resulting evolution of $u$ is shown in Fig. \ref{figepi}. Because of the Crank-Nicholson time integration, the amplitude remains constant up to round-off error. The phase accuracy is determined by the time step \citep{stone10}. For the case of Fig. \ref{figepi}, $\Omega \Delta t = 0.13$, which leads to a phase error of less than 1\% over the course of the simulation. 

\subsection{Linear shearing waves}
A more challenging problem, that also tests the self-gravity solver, consists of evolving a linear shearing wave \citep{gammie01}. Below, we first derive the governing equation, basically following \cite{gammie96}, but adapted to our notation and neglecting viscosity and magnetic fields.
 
\subsubsection{Governing equation}
Consider linear, adiabatic perturbations of the governing equations, so that the pressure perturbation $p_1 = \cs^2 \Sigma_1$, where $\cs$ is the adiabatic sound speed. The continuity equation and the two momentum equations then read
\begin{eqnarray}
\frac{1}{\Sigma_0}\frac{D\Sigma_1}{Dt}+\frac{\partial u_1}{\partial x} + \frac{\partial v_1}{\partial y}=0,\label{eqcontlin}\\
\frac{Du_1}{Dt}-2\Omega v_1 + \frac{1}{\Sigma_0}\frac{\partial p_1}{\partial x}+\frac{\partial \Phi_1}{\partial x}=0,\label{eqmomxlin}\\
\frac{Dv_1}{Dt}+(2-q)\Omega u_1 + \frac{1}{\Sigma_0}\frac{\partial p_1}{\partial y}+\frac{\partial \Phi_1}{\partial y}=0,\label{eqmomylin}
\end{eqnarray}
where $D/Dt = \partial/\partial t -q\Omega x \partial/\partial y$ is the convective derivative with respect to the unperturbed flow. Note that $D/Dt$ and $\partial/\partial y$ commute, but that
\begin{equation}
\frac{D}{Dt}\left(\frac{\partial f}{\partial x}\right)=\frac{\partial}{\partial x}\left(\frac{Df}{Dt}\right)+q\Omega \frac{\partial f}{\partial y}.
\end{equation}
Take the convective derivative of equation (\ref{eqcontlin}):
\begin{eqnarray}
\frac{1}{\Sigma_0}\frac{D^2\Sigma_1}{Dt^2}+\frac{D}{Dt}\left(\frac{\partial u_1}{\partial x}\right) + \frac{D}{Dt}\left(\frac{\partial v_1}{\partial y}\right)=\nonumber\\
\frac{1}{\Sigma_0}\frac{D^2\Sigma_1}{Dt^2}+\frac{\partial}{\partial x}\left(\frac{D u_1}{Dt}\right) +q\Omega \frac{\partial u_1}{\partial y}+\frac{\partial}{\partial y}\left(\frac{D v_1}{Dt}\right)=0,
\end{eqnarray}
and insert equations (\ref{eqmomxlin}) and (\ref{eqmomylin}):
\begin{eqnarray}
\frac{1}{\Sigma_0}\frac{D^2\Sigma_1}{Dt^2}+\frac{\partial}{\partial x}\left(2\Omega v_1 -\frac{1}{\Sigma_0}\frac{\partial p_1}{\partial x}-\frac{\partial \Phi_1}{\partial x}\right) + \nonumber\\
\frac{\partial}{\partial y}\left(2(q-1)\Omega u_1 - \frac{1}{\Sigma_0}\frac{\partial p_1}{\partial y}-\frac{\partial \Phi_1}{\partial y}\right)=0.\label{eqtemp}
\end{eqnarray}
The potential vorticity $\xi$ is given by 
\begin{equation}
\xi = \frac{(2-q)\Omega+\partial v/\partial x - \partial u / \partial y}{\Sigma},
\end{equation}
which can be approximated by a background value $\xi_0=(2-q)\Omega/\Sigma_0$ and a perturbation
\begin{equation}
\xi_1 = \frac{\partial v_1/\partial x - \partial u_1/\partial y}{\Sigma_0} -\frac{(2-q)\Omega}{\Sigma_0}\frac{\Sigma_1}{\Sigma_0}.
\end{equation}
For our test problem, we are interested in adiabatic perturbations, for which potential vorticity is conserved. Within the linear approximation, this means that $\xi$ is a shearing wave with constant amplitude, or $D\xi/Dt=0$. We must therefore have that
\begin{equation}
\frac{\partial u_1}{\partial y} = \frac{\partial v_1}{\partial x}-(2-q)\Omega \frac{\Sigma_1}{\Sigma_0}-\Sigma_0 \xi_1,
\label{eqvortpert}
\end{equation}
with $\xi_1$ a combination of shearing waves:
\begin{equation}
\xi_1 =\sum_{k_y} A_{k_y} (x) \exp(iq\Omega k_y t x+ik_y y).
\end{equation}
We can use equation (\ref{eqvortpert}) in equation (\ref{eqtemp}):
\begin{eqnarray}
\frac{1}{\Sigma_0}\frac{D^2\Sigma_1}{Dt^2}+\frac{\partial}{\partial x}\left(2q\Omega v_1 -\frac{1}{\Sigma_0}\frac{\partial p_1}{\partial x}-\frac{\partial \Phi_1}{\partial x}\right) - \nonumber\\
\frac{1}{\Sigma_0}\frac{\partial^2 p_1}{\partial y^2}-\frac{\partial^2 \Phi_1}{\partial y^2}-2(q-1)(2-q)\Omega^2\frac{\Sigma_1}{\Sigma_0}\nonumber\\
-2(q-1)\Omega\Sigma_0 \xi_1=0.\label{eqtemp2}
\end{eqnarray}
Differentiating the continuity equation with respect to $y$, and using equation (\ref{eqvortpert}), we obtain an expression for $v_1$:
\begin{eqnarray}
\left(\frac{\partial^2}{\partial x^2} + \frac{\partial^2}{\partial y^2}\right)v_1=(2-q)\Omega\frac{1}{\Sigma_0}\frac{\partial \Sigma_1}{\partial x}\nonumber\\
-\frac{1}{\Sigma_0} \frac{D}{Dt}\left(\frac{\partial \Sigma_1}{\partial y}\right)+\Sigma_0\frac{\partial \xi_1}{\partial x}.
\label{eqv1}
\end{eqnarray}
At this point, we decompose the solution into shearing waves:
\begin{equation}
X_1 = \hat X_1(t) \exp\left(i(k_x(t) x + k_y y)\right),
\end{equation}
with $k_x(t)=k_{x,0}+q\Omega k_y t$. For notational convenience, we set $\hat X_1=X_1$, the exponential factor being taken as read. Note that now $\xi_1$ is a constant. From equation (\ref{eqv1}) we obtain
\begin{equation}
k^2 v_1 = \frac{ik_y}{\Sigma_0} \frac{d\Sigma_1}{dt}-ik_x (2-q)\Omega \frac{\Sigma_1}{\Sigma_0}-ik_x \Sigma_0\xi_1,
\end{equation} 
with $k^2=k_x(t)^2+k_y^2$, while equation (\ref{eqtemp2}) reads
\begin{eqnarray}
\frac{1}{\Sigma_0}\frac{d^2\Sigma_1}{dt^2}+2q\Omega i k_x v_1 +k^2 \frac{p_1}{\Sigma_0} + k^2 \Phi_1-\nonumber \\ 
2(q-1)(2-q)\Omega^2\frac{\Sigma_1}{\Sigma_0}-2(q-1)\Omega\Sigma_0\xi_1=0.
\end{eqnarray}
Using the expression for $v_1$ above, $p_1 = \cs^2\Sigma_1$, and $\Phi_1 =  -2\pi G \Sigma_1/k$, we finally obtain
\begin{eqnarray}
\frac{1}{\Sigma_0}\frac{d^2\Sigma_1}{dt^2}-2q\Omega \frac{k_x k_y}{k^2}\frac{1}{\Sigma_0}\frac{d\Sigma_1}{dt}  +\nonumber \\
\left(2(2-q)\Omega^2 + k^2 \cs^2- 2\pi G\Sigma_0k -2q(2-q)\Omega^2\frac{k_y^2}{k^2}\right)\frac{\Sigma_1}{\Sigma_0}+\nonumber\\
2\Omega\left(1-q\frac{k_y^2}{k^2}\right)\Sigma_0\xi_1=0.
\label{eqlinsh}
\end{eqnarray}
This equation is equivalent to equation (15a) of \cite{gammie96}, but without viscosity and magnetic fields.

\subsubsection{Numerical results}
We consider a shearing wave in a sheet of size $L_x=L_y=1$ with $\Sigma_0=1/40$ and $Q=1$ of initial amplitude $\Sigma_1/\Sigma_0=0.0005$ with $k_x=-4\pi$ and $k_y=2\pi$. \cite{gammie01} considered a similar problem, but with $p_1=0$ initially, so that equation (\ref{eqlinsh}) does not apply since the initial perturbation is not adiabatic.

\begin{figure} 
\centering
\resizebox{\hsize}{!}{\includegraphics[]{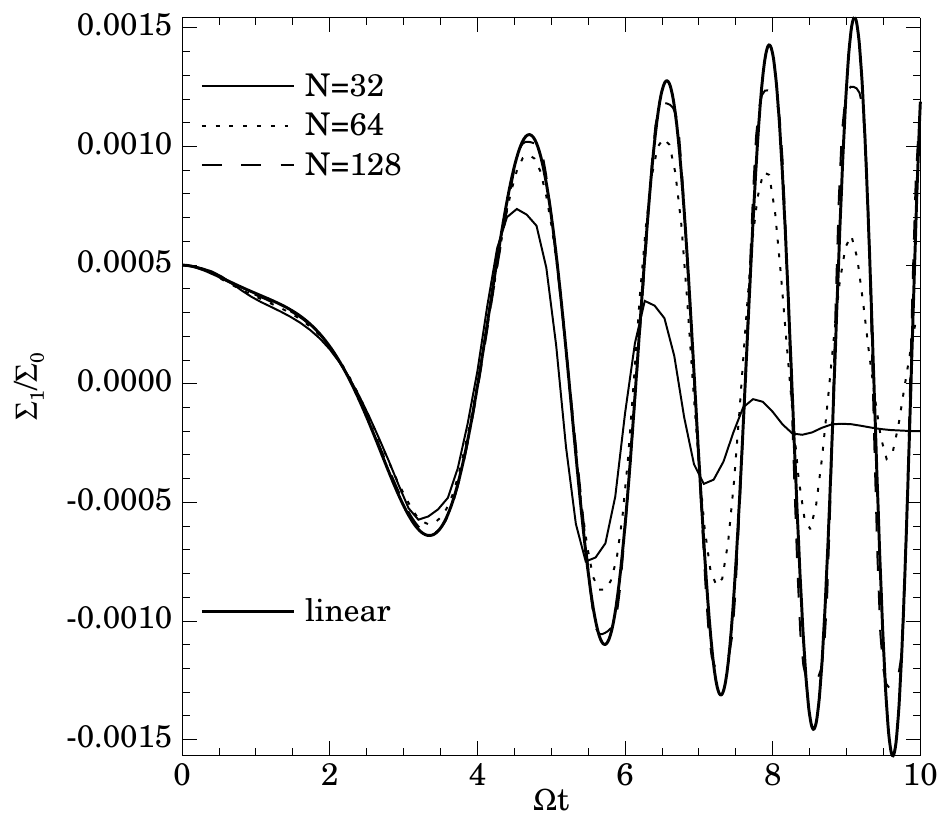}} 
\caption{Evolution of a non-selfgravitating linear shearing wave. The thick solid curve indicates the solution to equation (\ref{eqlinsh}), while the other curves indicate results from hydrodynamic simulations at different resolutions.} 
\label{figsh_lin} 
\end{figure} 

In Fig. \ref{figsh_lin}, we consider the case without self-gravity for different numerical resolutions. A fundamental scale to resolve is the scale height $H=\cs/\Omega\approx 0.07$. At the lowest resolution, $H$ is resolved by two grid cells only, which leads to strong diffusion of the wave. Since Riemann solvers actively use the sound speed, or, more general, the physical scales in the problem, to compute fluxes, not resolving the physical scale of the problem can lead to more excessive diffusion than for other numerical methods at comparable resolution. For eight grid cells per scale height ($N=128$), good agreement with linear theory is obtained up to $\Omega t=8$.

\begin{figure} 
\centering
\resizebox{\hsize}{!}{\includegraphics[]{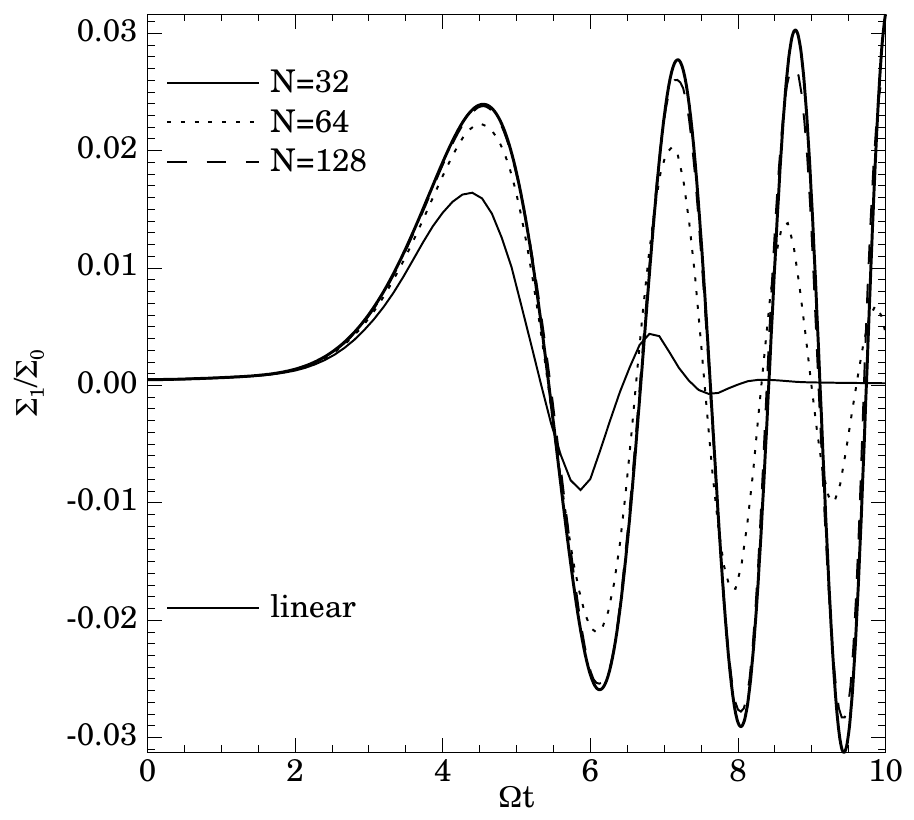}} 
\caption{Evolution of a self-gravitating linear shearing wave. The thick solid curve indicates the solution to equation (\ref{eqlinsh}), while the other curves indicate results from hydrodynamic simulations at different resolutions.} 
\label{figsh_lin_sg} 
\end{figure} 

We now turn to the case with self-gravity in Fig. \ref{figsh_lin_sg}, using $\delta=0$. For $N=128$, the numerical result follows linear theory in a similar way as for the case without self-gravity, which also compares well with \cite[][his Fig. 1]{gammie01}. Again, not resolving the physical scales of the problem leads to excessive diffusion for $N=32$, apparently more so than in \cite{gammie01}. This can be attributed to the fact that a Riemann solver actively uses the physical scale when computing the fluxes, so a penalty is paid if the necessary scales are under-resolved. 

\section{Initial conditions}
\label{secInit}
Initial conditions require special attention. In order for a steady gravito-turbulent state to be set up, we have to avoid initial transients \citep{conv}. Even though in local simulations no edges can form due to a radial dependence of the cooling time scale, it still takes a finite time for the turbulence to develop ($\Omega t_\mathrm{devel} \sim 10$). For cooling times smaller than or comparable to $t_\mathrm{devel}$, the disc will be cooled down to $Q<1$ before the disc can start to balance the cooling, which can trigger fragmentation even in cases where a steady gravito-turbulent state may exist. We therefore choose to hold $Q>1$ until $\Omega t = 50$. This is long enough for turbulence to develop and keep the disc from fragmenting artificially. After $\Omega t=50$, the disc is free to cool down and fragment. 

We take $\Sigma_0=1/320$ with $L_x=L_y=1$ and $Q=1$, similar to \cite{gammie01}. This makes $H=\cs/\Omega \approx  0.01$. The highest resolution considered by \cite{gammie01} was $N_x=N_y=1024$, resolving $H$ by approximately 10 grid cells. This will be our standard resolution, and we go up by factors of 2 from there. To compare with \cite{gammie01}, we use $\delta=0$ unless otherwise specified. The initial velocity field is seeded with white subsonic noise to let the turbulence develop. Following \cite{gammie01}, we take $\gamma=2$. We consider $16$ different cooling times, varying $\beta$ between $1$ and $50$. Each simulation is run until $\Omega t=1000$. Since we will find that fragmentation can be a stochastic process, we run four versions of a simulation, where we vary the phase of the initial seed noise, keeping the amplitude constant.

\section{Results}
\label{secRes}
The aim is to test for numerical convergence of the determination of the critical cooling time scale $\betac$. We do this by running simulations similar to those in \cite{gammie01}, but at higher resolution and for longer time spans. 

\subsection{Reproducing previous results}
\label{secPrev}
First of all, we try to reproduce previous results at the standard resolution ($N_x=N_y=1024$) and for $\Omega t < 100$.  As is usually done, we define the disc to have fragmented when an overdensity of $100\Sigma_0$ survives for several cooling time scales \citep[e.g.][]{meru11,rice11}.

We calculate the total stress in the sheet in the usual way \citep[e.g.][]{gammie01,rice11}. The average Reynolds stress is given by
\begin{equation}
\left<H_{xy}\right>=\left<\Sigma u v\right>,
\end{equation}
where $\left< \right>$ denotes an average over the whole computational domain. The gravitational stress is most easily determined in the Fourier domain:
\begin{equation}
\left<G_{xy}\right> = \sum_{\bf k} \frac{\pi G k_x k_y \left|\Sigma_{\bf k} \right|^2}{\left| {\bf k}\right|^3}.
\end{equation}
The total stress can be parametrised using the $\alpha$-prescription:
\begin{equation}
\alpha = \frac{2}{3}\frac{\left<H_{xy}\right> + \left< G_{xy}\right>}{\left<\Sigma \cs^2\right>},
\end{equation}
which can then be compared to equation (\ref{eqalpha}). We average the measured values of $\alpha$ over $\Omega \Delta t =20$ to get a single value for a given simulation. 

\begin{figure} 
\centering
\resizebox{\hsize}{!}{\includegraphics[]{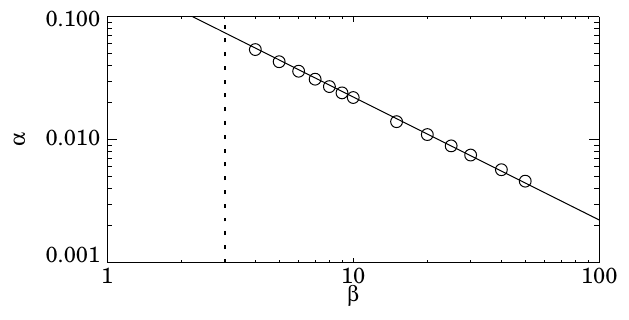}} 
\caption{Measured $\alpha$ parameter as a function of imposed cooling (open circles) together with the prediction of equation (\ref{eqalpha}) (solid line), for $N_x=N_y=1024$ and $\Omega t_\mathrm{max}=100$. The vertical dotted line shows the fragmentation boundary.} 
\label{fig_alpha_short} 
\end{figure} 
 
In Fig. \ref{fig_alpha_short}, we show the measured values of $\alpha$ together with the prediction of equation (\ref{eqalpha}). Over the range of $\beta$ we consider, we find agreement to within $5 \%$. Simulations with $\beta<4$ were found to fragment, which is in good agreement with \cite{gammie01}, who found $\betac=3$. Moreover, the maximum value of $\alpha$ the disc can sustain is $\alpha_\mathrm{max}=0.054$ (for $\beta=4$), in good agreement with \cite{rice05}, who found $\alpha_\mathrm{max}=0.06$. We also note that the measured rms density fluctuations agree with the results of \cite{cossins09}.

Since the physical scale of the instability, the most unstable wavelength $\lambda_T \sim H$ is resolved, one might argue that these results should be converged with respect to numerical resolution. Moreover, \cite{gammie01} showed that the measured value of $\alpha$ is independent of resolution if $N  \geq 512$. We have confirmed that the results depicted in Fig. \ref{fig_alpha_short} do not change when decreasing the resolution by a factor of 2. However, the value of $\alpha$ is not necessarily a good indicator of numerical convergence. Given the prescribed amount of cooling, the disc will try to generate enough heating to make up  for the energy that is removed. In the present set-up, it can only do that by generating the necessary stresses. Unless the simulation is dominated by numerical viscosity, the measured value of $\alpha$ will always be very close to the prediction of equation (\ref{eqalpha}); otherwise, the disc can not maintain a steady state. This, however, does not necessarily mean that the result makes sense, physically. In particular, convergence with respect to $\alpha$ does not imply convergence for the value of $\betac$.

In a similar way, resolving the physical scale of gravitational instabilities is only a necessary condition for numerical convergence. It would probably be sufficient if no other processes were going on, and if the evolution is predominantly on dynamical time scales. If very slow time scales on small scales are involved, it is likely that higher resolutions are required to capture the numerical evolution correctly.   We will see below that processes happening on the cooling time scale are critical in determining whether the disc will fragment or not. It is therefore expected that for increasing $\beta$, higher resolutions are required. 
    
\begin{figure} 
\centering
\resizebox{\hsize}{!}{\includegraphics[]{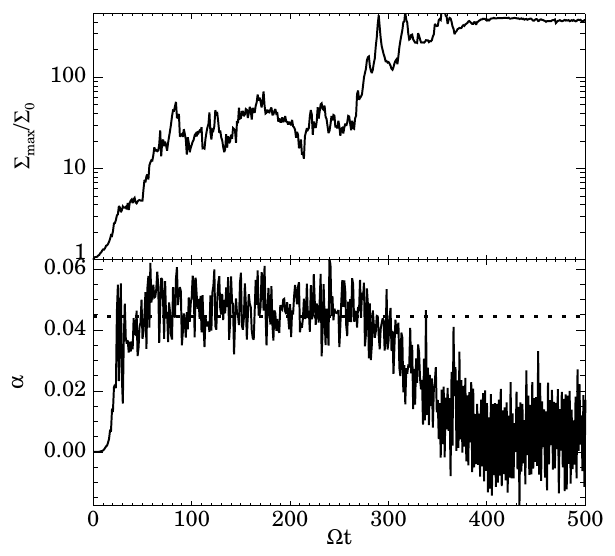}} 
\caption{Evolution of the maximum surface density (top panel) and the measured value of $\alpha$ (bottom panel) for a run with $N_x=N_y=1024$ and $\beta=5$. The dotted line in the bottom panel indicates the prediction of equation (\ref{eqalpha}).} 
\label{fig_1024_b5_r3} 
\end{figure} 

\begin{figure} 
\centering
\resizebox{\hsize}{!}{\includegraphics[]{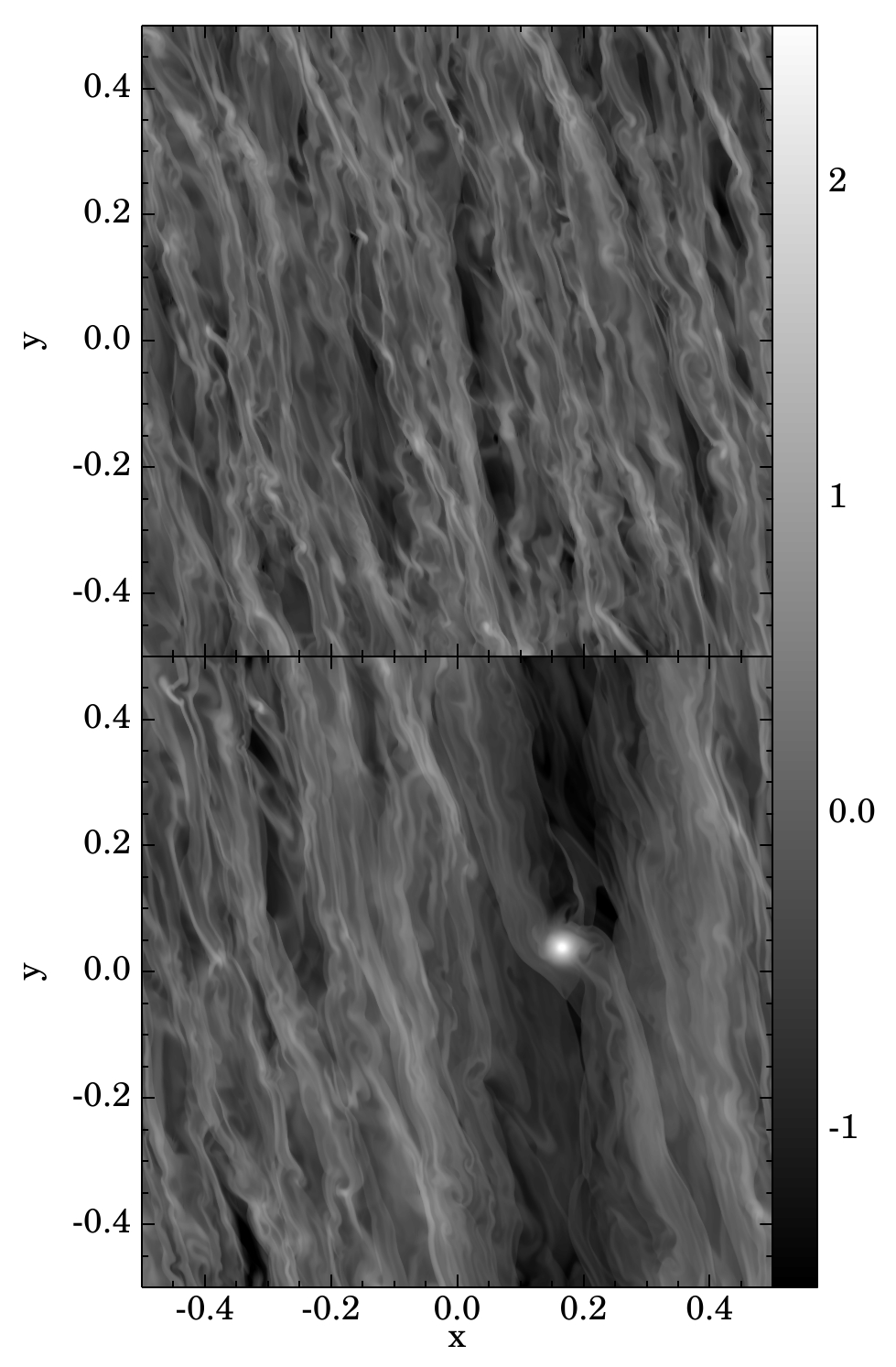}} 
\caption{Surface density, in terms of the initial surface density, on a logarithmic scale, for a simulation with $N_x=N_y=1024$ and $\beta=5$. Top panel: $\Omega t=100$, bottom panel: $\Omega t = 400$.} 
\label{fig_1024_b5_r3_t100_t400} 
\end{figure} 

\subsection{Longer time spans}
\label{secLong}
We now keep the resolution fixed at $N_x=N_y=1024$, but integrate all simulations until $\Omega   t_\mathrm{max}=1000$ (or until fragmentation occurs). In Figs. \ref{fig_1024_b5_r3} and \ref{fig_1024_b5_r3_t100_t400}, we focus on the case of $\beta=5$, which was found not to fragment for $\Omega t < 100$.

In the top panel of Fig. \ref{fig_1024_b5_r3}, the evolution of the maximum surface density in the sheet is shown. Before $\Omega t = 250$, $\Sigma_\mathrm{max}/\Sigma_0$ stays below 50, approximately, and the measured value of $\alpha$ agrees with equation (\ref{eqalpha}) (bottom panel of Fig. \ref{fig_1024_b5_r3}). The top panel of Fig. \ref{fig_1024_b5_r3_t100_t400} shows a snapshot of the surface density at $\Omega t = 100$. This looks like a very good example of steady gravito-turbulence, with density fluctuations that are consistent with those found by \cite{cossins09}.  

However, after $\Omega t=250$, something interesting happens. Suddenly, the maximum surface density shoots up to values above 100, indicating fragmentation. The bottom panel of Fig. \ref{fig_1024_b5_r3_t100_t400} shows a snapshot of the surface density at $\Omega t=400$, after the disc has fragmented. Only a single fragment was formed around $\Omega t = 250$, in contrast to simulations with $\beta < \betac$, which usually show $\sim 5-10$ fragments, initially, which can subsequently merge. 

The reason for this fragmentation at high values of $\beta$ lies in the nature of the gravito-turbulent state. Even before true fragmentation occurs, clumps are formed and destroyed on a continuous basis. This can be appreciated from the top panel of Fig. \ref{fig_1024_b5_r3}, where the peaks in $\Sigma_\mathrm{max}$ indicate a clump being destroyed. The root-mean-square density fluctuation is of order unity, while the maximum surface density reaches values of $\Sigma_\mathrm{max}/\Sigma_0=50$ several times. One clump that does not make it to collapse can be spotted near $x=0.05$ and $y=-0.45$ in the top panel of Fig. \ref{fig_1024_b5_r3_t100_t400}. 

Clumps of size $\sim H$ can survive the tidal shear if their size is less than the size of their Hill sphere. If we take the surface density within the clump to be constant for simplicity, we must have that
\begin{equation}
H<R_0 \left(\frac{\pi \Sigma H^2}{3 M_*}\right)^{1/3},
\label{eqCondShear}
\end{equation}
where $R_0$ is the radial distance to the central star and $M_*$ is its mass. This condition can be recast in terms of the local value of $Q$:
\begin{equation}
Q < \frac{1}{3}.
\end{equation}
In other words, keeping the temperature fixed, we only need an increase in surface density of a factor of $3$ over the background $Q_0\sim 1$. state to form a clump that can resist the shear. Once formed, these clumps will in general contract on a cooling time scale \citep{kratter11}. Their survival depends mainly on if they can resist the weak shocks that sweep around in gravito-turbulence. Since shock heating is very localised, this makes fragmentation a stochastic process: there will be a large spread in clump survival times, until the first lucky clump survives long enough for collapse to proceed. It should be noted that the condition given by equation (\ref{eqCondShear}) is not necessary if the cooling time scale is comparable to the dynamical time scale. If cooling acts on a dynamical time scale, there is no time for the clump to shear apart before it collapses. 

We have observed fragmentation up to $\beta=7$, more than twice the critical cooling time scale found by \cite{gammie01}. The corresponding maximum value of the stress is $\alpha_\mathrm{max}\approx 0.03$. For larger values of $\beta$, the disc remained in a steady, gravito-turbulent state for $\Omega t <1000$, with values of $\alpha$ that agree well with equation (\ref{eqalpha}).

\begin{figure} 
\centering
\resizebox{\hsize}{!}{\includegraphics[]{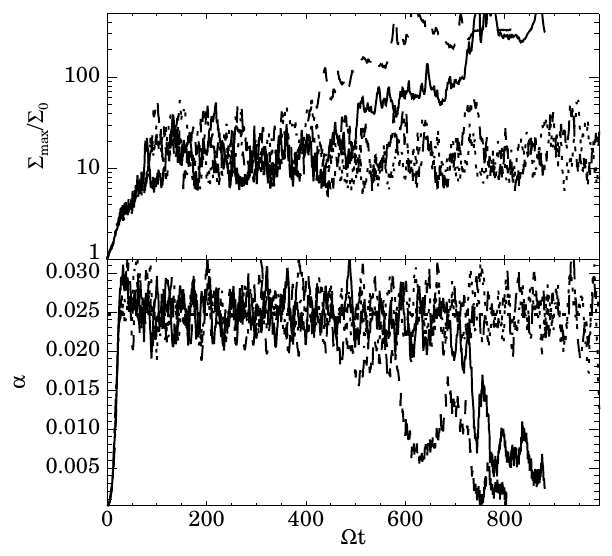}} 
\caption{Evolution of the maximum surface density (top panel) and the measured value of $\alpha$ (bottom panel) for four realisations with $N_x=N_y=2048$ and $\beta=9$. The dotted line in the bottom panel indicates the prediction of equation (\ref{eqalpha}).} 
\label{fig_1024_l_b9} 
\end{figure} 

\begin{figure} 
\centering
\resizebox{\hsize}{!}{\includegraphics[]{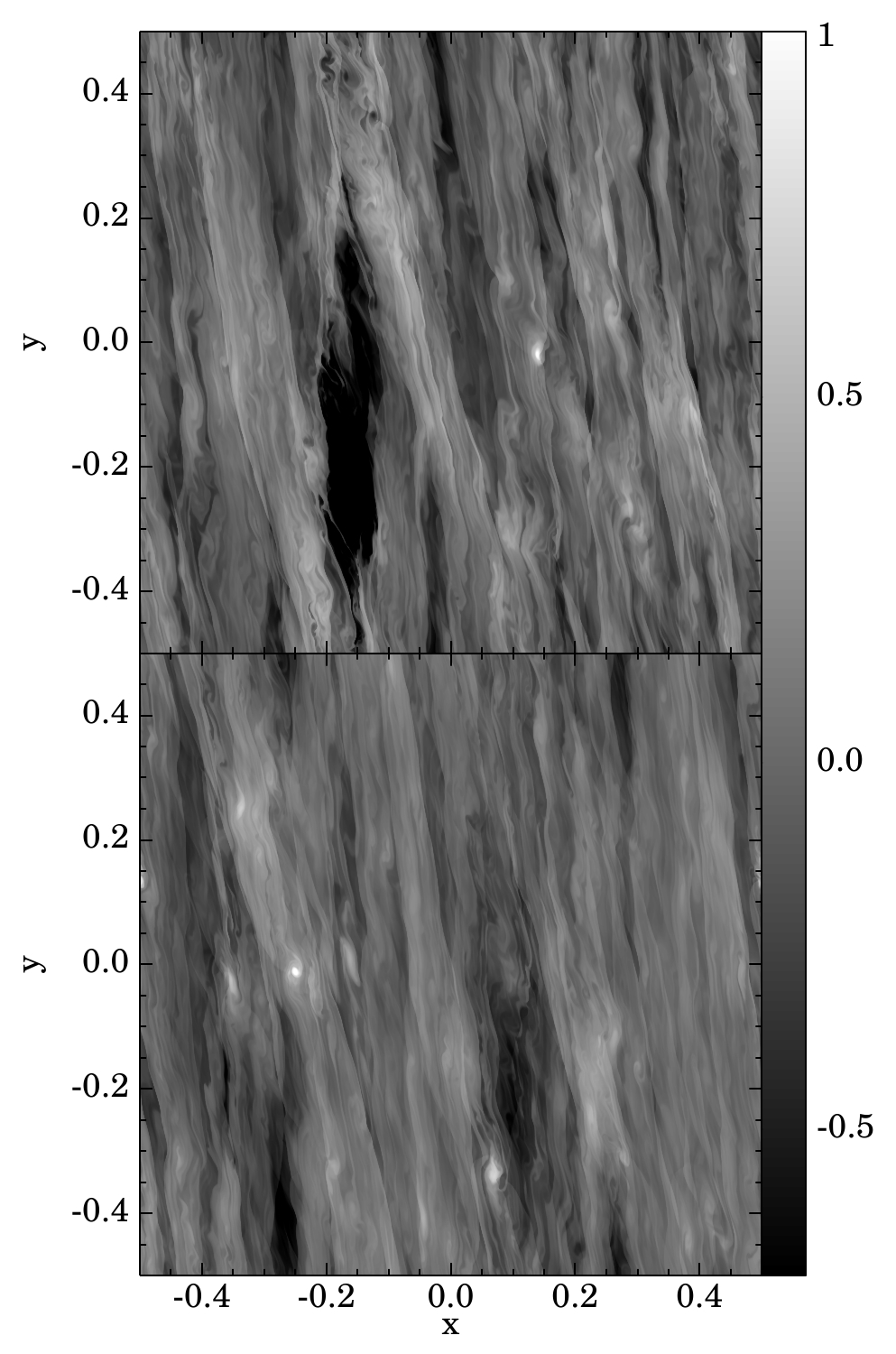}} 
\caption{Surface density, in terms of the initial surface density, on a logarithmic scale, for simulations with $N_x=N_y=2048$. Top panel: $\beta=20$, $\Omega t = 160$, bottom panel: $\beta=40$, $\Omega t = 370$.} 
\label{fig_2048_b20_b40} 
\end{figure} 

\subsection{Higher resolution}
We find that increasing the resolution by a factor of 2 ($N_x = N_y =2048$) leads to easier fragmentation at higher values of $\beta$. As an example, we show in Fig. \ref{fig_1024_l_b9} four simulations at $\beta=9$, differing only in the phase (not magnitude) of the initial noise. Two of the discs fragment, one at $\Omega t \approx 500$ and one at $\Omega t \approx 750$. The other two discs maintain a steady gravito-turbulent state for the full length of the simulation. This nicely illustrates the stochastic nature of disc fragmentation at high values of $\beta$: only in two out of four simulations does a clump survive for long enough for collapse to proceed. It is expected that if the simulations would be continued, in the end \emph{all} of the four realisations should show fragmentation. Note that fragmentation now occurs for three times higher values of $\beta$ than in Sect. \ref{secPrev}.

We have found clumps that can survive shear to form for all values of $\beta$ we have considered ($\beta \leq 50$). In Fig. \ref{fig_2048_b20_b40}, two examples are shown of clumps with $\Sigma/\Sigma_0 \approx 10$ for $\beta=20$ (top panel) and $\beta=40$ (bottom panel). However, while clumps form readily in all simulations, the vast majority do not survive. In the top panel of Fig. \ref{fig_1024_l_b9}, at least 10 clumps were formed and destroyed before the first disc fragments. In discs that do not fragment, more than 20 clumps form during the time span of the simulation, but none of them collapse into bound fragments. Given this low success rate, simulations should span at least 100 cooling time scales to capture these events. This becomes impractical for very high values of $\beta$.  The highest value of $\beta$ for which we have found fragmentation is $\beta=20$ (with $N_x=N_y=4096$), almost 7 times higher than what was found in Sect. \ref{secPrev}. 

The reason why this behaviour can not be captured at low resolution, lies in the fact that we need the clumps to survive for $\Omega \Delta t \sim \beta$. It is not enough just to resolve the length scale $H$, the numerical scheme needs to be able to maintain a coherent clump of size $H$ over many dynamical time scales. The resolution required will depend on the details of the numerical implementation, and will increase for larger values of $\beta$. In addition, long time integrations are needed to weed through all the clumps that fail to collapse.  

\begin{figure} 
\centering
\resizebox{\hsize}{!}{\includegraphics[]{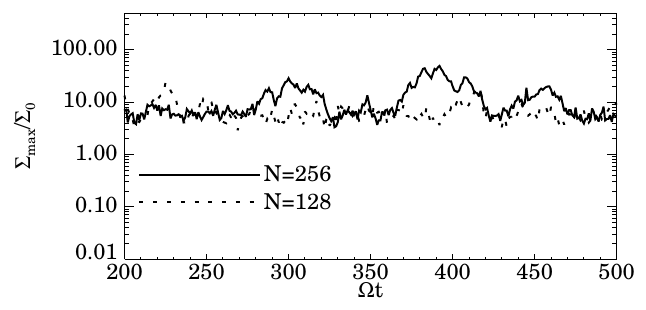}} 
\caption{Evolution of the maximum surface density for simulations with $\beta=8$ in a domain that is reduced in size by a factor of $8$, for two different resolutions.} 
\label{fig_smallbox} 
\end{figure} 

The need for high resolution can be appreciated further by looking at smaller domains, so that only a single transient clump is present at any time. The evolution of the maximum surface density is then linked directly to the evolution of this clump. In Fig. \ref{fig_smallbox}, we show the evolution of the maximum surface density for $\beta=8$ in a domain that is reduced by a factor of $8$ compared to the standard run (i.e. $L_x=L_y=1/8$). Note that still $H \ll L_x$, and that $N=128$ now corresponds to our standard resolution. Neither of the two simulations shows fragmentation, but transient clumps can be clearly identified. In the high-resolution run, it takes approximately $\Omega t=40$ for a clump to reach its maximum density, which corresponds to 5 cooling time scales. In the low resolution run, clumps do not survive this long, leading to less-pronounced surface density peaks. Therefore, clump survival appears to be linked to numerical resolution.  

The frequency at which clumps appear decreases for higher values of $\beta$. While for $\beta=9$, approximately $20$ failed clumps can be identified over a time span of $\Omega t = 1000$, for $\beta=50$ only $\sim 4$ failed clumps appear. The frequency appears to be roughly proportional to $1/\beta$. Together with the fact that the clumps contract on a cooling time scale, this makes fragmentation very rare, but not impossible, for higher values of $\beta$. 

\begin{figure} 
\centering
\resizebox{\hsize}{!}{\includegraphics[]{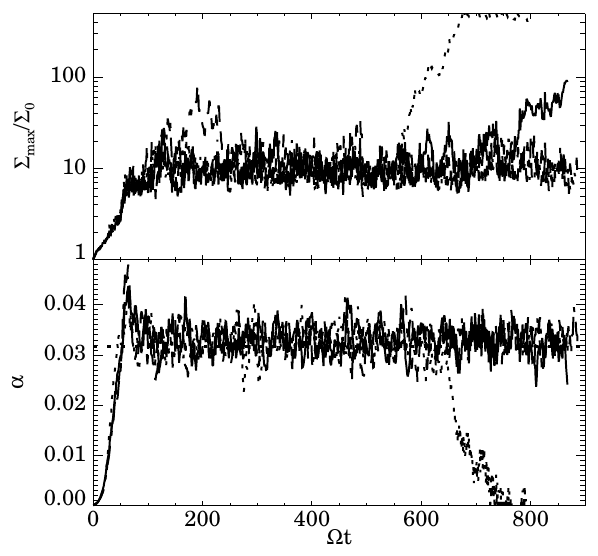}} 
\caption{Evolution of the maximum surface density (top panel) and the measured value of $\alpha$ (bottom panel) for four realisations with $N_x=N_y=2048$ and $\beta=7$ with $\delta=H$. The dotted line in the bottom panel indicates the prediction of equation (\ref{eqalpha}).}
\label{fig_soft} 
\end{figure} 

\subsection{Additional numerical effects}
\label{secsmooth}
In this section, we discuss the influence of two more numerical parameters: the smoothing length $\delta$ and the flux limiter $\phi$.  Previous studies \citep{gammie01,rice11} used $\delta=0$, which, at high resolution, allows for variations in the potential on scales much smaller than a scale height $H$. If we interpret the two-dimensional approximation as resulting from vertically averaging the three-dimensional equations, we do not expect to find such small scales. If these scales are important, this calls for fully three-dimensional simulations. We have performed additional runs using $\delta = \cs/\Omega$, where the average sound speed was calculated every time step. Note that this is a fairly large value for $\delta$; in two-dimensional disc-planet interaction studies, the gravitational potential of the planet is usually smoothed over a distance $H/2$ \citep[e.g.][]{baruteau08}. Moreover, the true disc thickness will be smaller than $\cs/\Omega$ because of self-gravity. Using $\delta=\cs/\Omega$ may therefore introduce more smoothing than necessary. 

The results for $\beta=7$ are shown in Fig. \ref{fig_soft}. As in the case with $\delta=0$, fragmentation appears to be stochastic, with one out of four realisations forming a bound fragment within $\Omega t=1000$. This effect therefore does not rely heavily on small scales in the gravitational potential. However, no fragmentation was found for $\beta>10$, in contrast to the simulations with $\delta=0$. Thus, while the overall picture is very similar, with transient clumps clearly visible in Fig. \ref{fig_soft}, the efficiency of stochastic fragmentation is reduced for $\delta=\cs/\Omega$. Three-dimensional simulations are necessary to see which case is appropriate. 

A similar story holds for changing the flux limiter. Simulations using the diffusive minmod flux limiter, obtained by setting $s=1$ in equation (\ref{eqfluxlimiter}), still show transient clumps and stochastic fragmentation, but only up to $\beta = 5$ for the standard resolution. This is expected: since the minmod limiter introduces more numerical diffusion, it is more difficult for transient clumps to survive long enough to collapse. It is worth pointing out that in some multi-dimensional problems, the minmod limiter appears to give more diffusion than finite difference codes \citep{bingas}.

\section{Discussion}
\label{secDisc}
We have shown in the previous section that disc fragmentation is a stochastic process in the simplified system under consideration. It is likely that some of the simplifications made will affect the efficiency of disc fragmentation for longer cooling times. We discuss the three most important ones below, before we focus on possible implications. 

\subsection{Limitations}
First of all, we have worked in two dimensions only. This means that all quantities should be thought of as being integrated over the disc scale height $H$. While this is straightforward for the gas density, velocities and pressure, the potential due to self-gravity poses a problem. The force due to self-gravity should be smoothed over a length comparable to the scale height. While introducing a smoothing length did not change the results in a qualitative way (see Sect. \ref{secsmooth}), three-dimensional simulations are needed to determine whether the smoothed or the unsmoothed results are more appropriate for three-dimensional discs.

Second, the use of the simple cooling law of equation (\ref{eqtcool}) is questionable as soon as fragments form. When the density goes up by orders of magnitude, cooling should slow down dramatically. This can be a very important effect in the present situation, where clumps are formed on a continuous basis, trying to cool down before they are destroyed. The survival of clumps strongly depends on their ability to cool, and it is likely that when cooling slows down at increasing surface density, the likelihood of survival goes down.  

Finally, we have considered \emph{local} models only. While this is advantageous for studying a steady gravito-turbulent state, it is impossible to say how the disc came into this state \citep[e.g.][]{kratter11} or what happens to any fragments that have formed. For example, it is likely that these newly-formed objects are subject to rapid radial migration \citep{baruteau11}. Also, any interaction with global modes \citep{lodato05} is necessarily excluded from the current study.

\subsection{Implications}
The most important conclusion from the results presented in Sect. \ref{secRes} is that there is no such thing as a steady gravito-turbulent state in the simple case of $\beta$-cooling, at least not for $\beta<20$. Even though the disc can balance the imposed cooling for many dynamical time scales (see Fig. \ref{fig_1024_l_b9}), there is always a chance the disc will fragment at some point. This means there is no rock-solid criterion for disc fragmentation based on the cooling time scale. Fragmentation just gets less likely for higher values of $\beta$. The classical cooling criterion, which states that cooling should occur on a dynamical time scale, is in a way a condition for fragment survival as well: for $\beta \sim 1$, clumps can not be sheared apart by tidal forces (see Sect. \ref{secLong}), ensuring essentially a $100$ \% survival rate.  

The maximum time span we have considered is $\Omega t_\mathrm{max} =1000$. Since the self-gravitating phase of protoplanetary discs is thought to last only for $\sim 10^5$ years \citep[e.g.][]{laughlin94}, if we take our sheet to be located at 100 AU, there is less than $\Omega t_\mathrm{max}$ available before the disc is no longer self-gravitating. It has to be kept in mind that even though steady gravito-turbulence may not exist formally in the simple case of $\beta$-cooling, it is only necessary to be steady for a finite time span. There is therefore a value of $\beta$ for which fragmentation becomes impractical, because it just takes too long for a rare event (the survival of a clump) to happen. It is difficult to say, for a given value of $\beta$, exactly when the disc will fragment. This will for example depend on the total area of the sheet under consideration; larger sheets have a better chance to fragment. Since the vast majority of clumps will not survive, and since they contract on a cooling time scale, it is necessary to integrate for several 100's of cooling time scales to see fragmentation.    
 
Based on these results for discs with simplified thermodynamics, it is dangerous to model the self-gravitating phase of disc evolution using an $\alpha$-model. As soon as fragmentation sets in, stresses become dominated by the fragment (see for example the bottom panel of Fig. \ref{fig_1024_b5_r3}). The extent to which the fragment can come to dominate will depend on its final mass, which is likely not to be captured very well in the current simulations because of the simple cooling law. If cooling becomes less efficient at higher density, the growth time of the fragments will likely go up, and the impact of the fragment will not be as dramatic as depicted in Fig. \ref{fig_1024_b5_r3}. Moreover, it is likely that the fragment will be subject to rapid orbital migration \citep{baruteau11}, leaving its place of birth and perhaps allowing the disc to resettle into a gravito-turbulent state, albeit at lower mass. However, this all depends very sensitively on the mass evolution of the fragments, which is poorly constrained at the moment \citep[see e.g.][]{kratter10, boley10}.

Fragmentation at higher values of $\beta$ does not necessarily make planet formation by GI possible in the inner regions of protoplanetary discs, since it is likely that $\beta$ increases very rapidly towards the central star. The cooling time scale is proportional to \citep[see][]{kratter10}
\begin{equation}
t_\mathrm{cool} \propto \frac{\Sigma^2 \cs^2 \kappa_\tau}{T^4},
\end{equation}
where $\kappa_\tau$ is the opacity per unit mass. For an opacity law $\kappa_\tau \propto \Sigma^a T^b$, we get, using the ideal gas law:
\begin{equation}
t_\mathrm{cool} \propto \Sigma^{2+a} \cs^{2b-6}.
\end{equation}
In a gravito-turbulent state, $Q$ is constant, so we must have that $\Sigma \propto \cs\Omega$, and therefore:
\begin{equation}
\beta = t_\mathrm{cool}\Omega \propto \cs^{a+2b-4} \Omega^{3+a}.
\end{equation}
In the outer regions of protoplanetary discs, we expect $a=0$ and $b=2$ \citep[see][]{bell94}, which leads to $\beta \propto \Omega^3 \propto R_0^{-9/2}$. Since $\beta$ is such a steep function of radius, it is very difficult to fragment at short distances, even in view of the results of this paper. However, because of possible inward migration, there is no need to fragment closer in, provided that the fragments can stay in the planetary mass regime. In principle, the mass of a fragment can even decrease if it migrates inward because of tidal stripping \citep{boley10,nayakshin10}.

Another way of saying that there is no critical cooling time scale for fragmentation in the simple case of $\beta$-cooling studied here, is that there is no well-defined maximum stress the disc can sustain \citep{rice05}. Formally, $\alpha_\mathrm{max}=0$. In practice, however, fragmentation becomes very rare for high values of $\beta$, but in general great care has to be taken in using either $\betac$ or $\alpha_\mathrm{max}$. There is no sharp boundary between discs that show fragmentation and discs that do not. 

\section{Conclusions}
\label{secCon}
In this paper, we have studied the numerical convergence of the determination of the critical cooling time scale for disc fragmentation. We have seen that, in the two-dimensional local approximation with a simple cooling law, there is no sharp boundary in terms of the cooling time scale between discs that fragment and discs that do not. A `steady', gravito-turbulent state consists of weak shocks as well as transient clumps, that will contract on a cooling time scale. If such a transient clump can survive in the turbulent background for long enough, it will collapse and form a bound fragment. Since the weak shocks affect the disc only very locally, it is possible in principle for clumps to survive for many dynamical time scales. This makes disc fragmentation a stochastic process: most of these transient structures will be destroyed, but in the end one lucky clump will make it into a fragment. Transient clumps were found for all cooling times considered, and therefore fragmentation is possible in principle for cooling times up to $\beta=50$. However, fragmentation becomes increasingly rare for longer cooling time scales. 

It is possible that the efficiency of stochastic disc fragmentation is affected by the approximations used. It remains to be seen whether similar effects can be observed in three-dimensional simulations with a more realistic cooling prescription.

\section*{Acknowledgements}
I would like to thank the anonymous referee, whose constructive comments led to an improvement of the paper.  SJP is supported by an STFC postdoctoral fellowship.  Simulations were performed using the Darwin Supercomputer of the University of Cambridge High Performance Computing Service (http://www.hpc.cam.ac.uk), provided by Dell Inc. using Strategic Research Infrastructure Funding from the Higher Education Funding Council for England.

\bibliography{paardekooper}

\label{lastpage}

\end{document}